\documentclass[12pt]{iopart}

\usepackage[pdftex]{graphicx}

\usepackage{cite}

\usepackage{color} 

\date{}

\newcommand{\be}{\begin{eqnarray}}
\newcommand{\ee}{\end{eqnarray}}

\begin{document}

\title{Punishment in Public Goods games leads to meta-stable phase transitions and hysteresis}
\author{Arend Hintze$^{1,2}$}
\author{Christoph Adami$^{1,2,3}$}
\address{$^{1}$Department of Microbiology and Molecular Genetics, Michigan State University, East Lansing, MI}
\address{$^{2}$BEACON Center for the Study of Evolution in Action, Michigan State University, East Lansing, MI}
\address{$^{3}$Department of Physics and Astronomy, Michigan State University, East Lansing, MI}
\ead{adami@msu.edu}

\date{} 
\begin{abstract}The evolution of cooperation has been a perennial problem in evolutionary biology because cooperation can be undermined by selfish cheaters who gain an advantage in the short run, while compromising the long-term viability of the population. Evolutionary game theory has shown that under certain conditions, cooperation nonetheless evolves stably, for example if players have the opportunity to punish cheaters that benefit from a public good yet refuse to pay into the common pool. However, punishment has remained enigmatic because it is costly, and difficult to maintain. On the other hand, cooperation emerges naturally 
in the Public Goods game if the synergy of the public good (the factor multiplying the public good investment) is sufficiently high. In terms of this synergy parameter, the transition from defection to cooperation can be viewed as a  phase transition with the synergy as the critical parameter. We show here that punishment reduces the critical value at which cooperation occurs, but also creates the possibility of meta-stable phase transitions, where populations can ``tunnel" into the cooperating phase below the critical value. At the same time, cooperating populations are unstable even above the critical value, because a group of defectors that are large enough can ``nucleate" such a transition. We study the mean-field theoretical predictions via agent-based simulations of finite populations using an evolutionary approach where the decisions to cooperate or to punish are encoded genetically in terms of evolvable probabilities. We recover the theoretical predictions and demonstrate that the population shows hysteresis, as expected in systems that exhibit super-heating and super-cooling. We conclude that punishment can stabilize populations of cooperators below the critical point, but it is a two-edged sword: it can also stabilize defectors above the critical point.
 
\end{abstract}
\section{Introduction}
When individuals maximize their self-interest by exploiting a public good, they are often doing so by harming their (and others') own long-term interest, and create a social dilemma termed the ``tragedy of the commons"~\cite{Hardin1968}. The tragedy of the commons is often discussed in environmental politics (for example, overgrazing and overfishing), as well as social science and politics (for example, vandalism and taxation)~\cite{Hardin1968}. However, the tragedy of the commons also plays an important role in evolutionary biology~\cite{Rankinetal2007}: rate-yield tradeoffs in bacterial metabolism~\cite{MacLean2008}, the evolution of virulence~\cite{Frank1996} and the manipulation of a host by a group of parasites~\cite{Brown1999} can be viewed as a social dilemma involving a public good. 
Social dilemmas~\cite{Frank2006} (such as the tragedy of the commons) can be studied within the framework of Evolutionary Game Theory (EGT)~\cite{MaynardSmith1982,Axelrod1984,Dugatkin1997,HofbauerSigmund1998,DoebeliHauert2005,Nowak2006}, which describes populations of agents engaging in pairwise (or group wise) interactions, with defined payoffs for different strategies. The tragedy of the commons is usually described by a particular game form known as the  ``Public Goods" game. 


The Public Goods game is a standard within the field of experimental economics~\cite{Olson1971,DavisHolt1993,Ledyard1995}. In this game, players possess tokens that they can invest into a common pool (the public good). The total sum contributed by the players is then multiplied by a ``synergy factor" (creating a positive yield). This amount (typically larger than the invested sum) is then equally distributed to the players in the pool, irrespective of whether they invested or not. A group of players maximizes their investment if all the players contribute (so as to take maximum advantage of the synergy). However, this behavior is vulnerable to ``free-riders": individuals that share in the pool but do not invest themselves. It can easily be shown that the rational Nash equilibrium for this game is {\em not} to pay in, because this strategy clearly dominates all others regardless of their play~\cite{Hardin1968}.

Hardin originally suggested that the tragedy of the commons can only be avoided by {\em punishing} free riders~\cite{Hardin1968}. Indeed, it has been shown that punishment can counteract defectors effectively~\cite{FehrGachter2002,FehrFischbacher2003,Hammerstein2003,NakamaruIwasa2006,CamererFehr2006,Gurerketal2006,Sigmundetal2001,HenrichBoyd2001,Boydetal2003,Brandtetal2003,Helbingetal2010}, but punishment is difficult to maintain because it is costly and may reduce the mean payoffs of group members compared to groups in which punishment of free-riders is not possible~\cite{Boydetal2010}. 

It was previously thought that punishment cannot be maintained in well-mixed populations~\cite{Helbingetal2010c} so most of the literature has focused on the spatial version of the game, where analytical results are difficult (but not impossible) to obtain~\cite{SzaboHauert2002}. 
More recent work has uncovered a decidedly more complicated picture, in particular because punishment can be performed in a number of different ways, that each change the dynamics of the population considerably. For example, if the punisher also rewards cooperators at the same time~\cite{SzolnokiPerc2013b}, complex dynamics that depend on the strength of punishment and the size of the synergy can emerge in the spatial game, however the reward/punish strategy is only stable in rare circumstances. It is also possible to punish in such a manner that the severity of punishment depends on the number of defectors~\cite{PercSzolnoki2012}. In the spatial version of the game, introducing such an ``adaptive" punishment leads to a strong enhancement of the cooperative phase, as rare punishers protect the boundary between the phases with maximum punishment. Another variation of punishment is to introduce a conditional punisher, that scales punishment with the number of defectors in the group~\cite{SzolnokiPerc2013a}. In that game (also in the spatial setting), conditional and unconditional punishers cannot coexist, and who ultimately wins depends on the strength of punishment. There are variations of the Public Goods Game in which punishment can be maintained in other ways, for example by voluntary punishment~\cite{SzaboHauert2002,Fowler2005,Hauertetal2007,DeSilvaetal2009,Sigmundetal2010} or by using pool (that, is institutionalized) as opposed to peer punishment~\cite{Yamagishi1986,Sigmundetal2010,Sasakietal2011,Szolnokietal2011}, but we do not study those here.

A variant of the game that is closer to the one we study here involves {\em probabilistic} punishment, where rather than punishing any defector with certainty, cooperators punish with probability $\pi$. Chen et al.~\cite{Chenetal2014} show that introducing a probability to punish changes the fixed point structure of the game in the well-mixed regime so that a repulsive interior fixed point can appear, and this enhances cooperation also in the spatial version of the game. 

Here we study the well-mixed version of the Public Goods game with probabilistic punishment and probabilistic cooperation, establish a number of theoretical results that suggest complex dynamics at the interface between the cooperating and defecting phases (as a function of the synergy factor) and clarify the role of punishment as a catalyst of cooperation in extensive agent-based simulations that agree with the theoretical results.

\section{Mean field theory of Public Goods games} \label{sec-theory}
The Public Goods game emulates strategic decision making by groups, in which an individual must select between different decisions that affect the group as a whole.
Each individual in a group of $k+1$ players (the focal player and her $k$ participants) can decide to cooperate by making a contribution of 1 unit to the public good, while defecting individuals do not contribute. 

\begin{figure}[htbp] 
   \centering
   \includegraphics[width=3.5in]{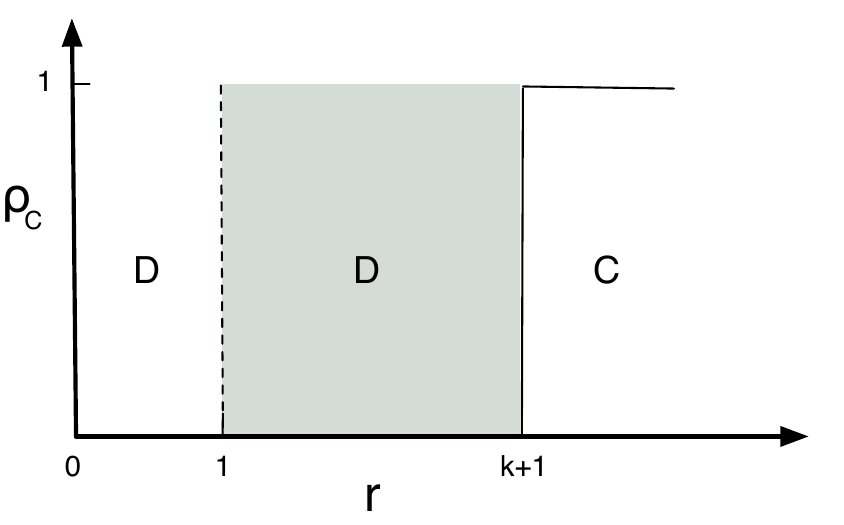} 
   \caption{The phase diagram of the Public Goods game with a synergy factor $r$. Below $r=1$ defection is the strategy with the highest payoff and therefore favored by evolution. Conversely, above $r=k+1$ cooperation is evolutionarily favored. A dilemma exists in the grey-shaded area between $r=1$ and $r=k+1$, where cooperation would be beneficial for a cooperating group as a whole, but defection is the Nash equilibrium point and thus evolutionarily favored. }
   \label{dilemma}
\end{figure}

The sum of all contributions from cooperating players is multiplied by $r$ (the synergy factor) and divided among all players. If $N_C$ is the number of cooperators within the group (but not counting the focal player, i.e., $N_C\leq k$) and $N_D$ is the number of defectors, then
the  cooperator obtains a payoff
\be
P_C = r\frac{(N_C+1)}{k+1} -1 \label{eq1a}
\ee
compared to the defector's
\be
P_D=r\frac{N_C}{k+1}\;. \label{eq2a}
\ee
A dilemma exists if it is advantageous for the individual to defect, while mutual cooperation would be best for all. Clearly a defector does better if $P_D-P_C>0$, so a dilemma exists only if $r<k+1$. At the same time, the payoff for a cooperator playing within a group of cooperators should be larger than the payoff for a defector playing only with defectors, that is, $P_C(N_C=k)-P_D(N_C=0)>0$ which implies $r>1$.
Thus, a dilemma exists only for $1<r<k+1$ (see Fig.~\ref{dilemma}). 
Standard evolutionary game theory arguments imply that defection is the rational (and optimal) strategy for $r<k+1$, while cooperation is selected for when $r>k+1$. The synergy parameter $r$ can thus be viewed as a critical parameter, dialing a phase transition from defection to cooperation as $r$ is increased through $k+1$. This transition is (in the limit of infinite population size and vanishing mutation rate) of first-order (meaning an abrupt discontinuous transition from defection to cooperation as the synergy parameter is raised across its critical value), as can be seen from the replicator equation for the density of cooperators $\rho_C$
\be
\dot\rho_C=\rho_C(1-\rho_C)(P_C-P_D)\;.
\ee
Note that the Public Goods game turns into the standard Prisoner's dilemma for $k=1$, with a dilemma for $1<r<2$. 

How can this dilemma be solved? How can evolution achieve cooperation in the grey-shaded area in Fig.~\ref{dilemma}? The answer is: this is impossible unless additional mechanisms change the critical point below $r_c=k+1$. One such mechanism investigated in the literature is giving players the option to punish players who do not contribute. 
Following the notation of Helbing et al.~\cite{Helbingetal2010}, defecting players suffer a {\em fine} $\beta/k$ levied by each punisher in the group, which costs each punisher a {\em penalty} $\gamma/k$. Let $N_M$ be the number of players that cooperate as well as punish (the ``moralists") and $N_I$ the number of defectors that punish (``immoralists"). As before, $N_C$ and $N_D$ are the number of players that cooperate viz. defect but do not punish. The payoffs for the four possible strategies then become 
\be
P_C &=& r\frac{(N_C+N_M+1)}{k+1} -1\;, \label{eq1}\\
P_D&=&r\frac{(N_C+N_M)}{k+1}-\beta\frac{(N_D+N_I)}{k}\;, \label{eq2}\\
P_M&=&P_C-\gamma\frac{(N_D+N_I)}{k}\;,\\
P_I&=&P_D-\gamma\frac{(N_D+N_I)}{k}\;.
\ee

\subsection{Meean-field theory}
\begin{figure}[htbp] 
   \centering
   \includegraphics[width=4in]{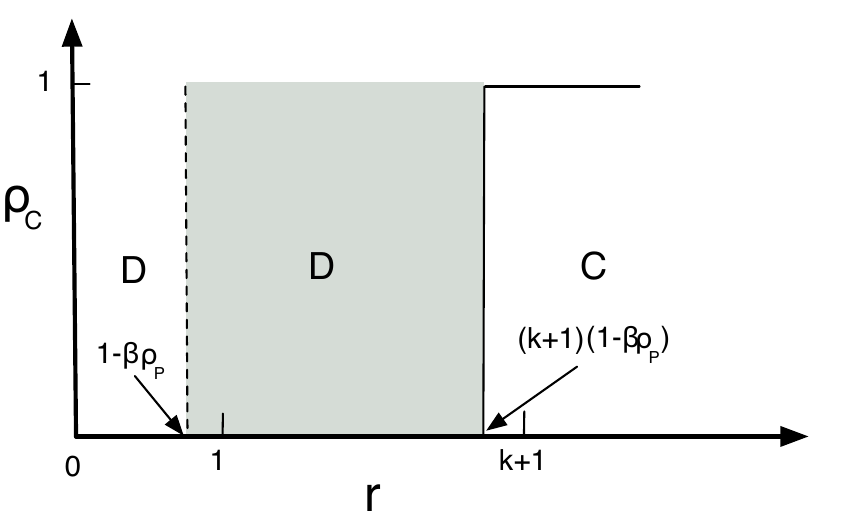} 
   \caption{The phase diagram of the Public Goods game with a synergy factor $r$ and punishment. Compared to Fig.~\ref{dilemma}, the area where a dilemma occurs is shifted towards lower $r$, implying that cooperation can occur for smaller $r$ (right boundary of dilemma area). A single defector cannot invade the population to the right of the critical point, and a single cooperator cannot invade to the left of $r_c$ }
   \label{dilemma2}
\end{figure}

Let us calculate the critical point for the game with punishment, assuming a well-mixed population so that each player encounters on average the same fraction of strategies. Introducing the mean density of cooperators $\rho_C$ and the mean density of punishers $\rho_P$
\be
\rho_C&=&\frac{N_C+N_M}k \label{rhoc}\\
\rho_P&=&\frac{N_M+N_I}k \label{rhop}\;,
\ee
along with $\rho_D=\frac{N_D+N_I}k $ we can write the average payoffs for each of the four strategies as
\be
P_C &=& r\frac{(k\rho_C+1)}{k+1} -1\;, \label{eq1c} \\
P_D&=&r\frac{k\rho_C}{k+1}-\beta \rho_P\;,\label{eq2c} \\
P_M&=&P_C-\gamma\rho_D\;, \label{eq3c}\\
P_I&=&P_D-\gamma\rho_D\;.\label{eq4c}
\ee
Investigating $P_C-P_D$ again, we notice that the area where the dilemma exists is now shifted by $\beta\rho_P$ (see also Fig.~\ref{dilemma2}): 
\be
1-\beta\rho_P<r<(k+1)(1-\beta\rho_P)\;.
\ee
where the right boundary corresponds to the critical point $r_c=(k+1)(1-\beta\rho_P)$ separating a cooperating and a defecting phase.  Because we will be concerned with this critical point from now on, let us introduce the re-scaled synergy parameter $\xi=r/(k+1)$. The critical point is then $\xi_c=1-\beta\rho_P$. 
According to standard population genetics, a single cooperating individual cannot invade a population of defectors unless its fitness advantage $P_C-P_D$ is positive, which implies $\xi>1-\beta\rho_P$. However, if the entire population consists of defectors, punishment is expected to be absent because defectors do not punish each other (i.e., we assume here that immoralists do not matter in the long run, as was found in numerical simulations~\cite{Helbingetal2010}). 

\begin{figure}[htbp] 
   \centering
   \includegraphics[width=4in]{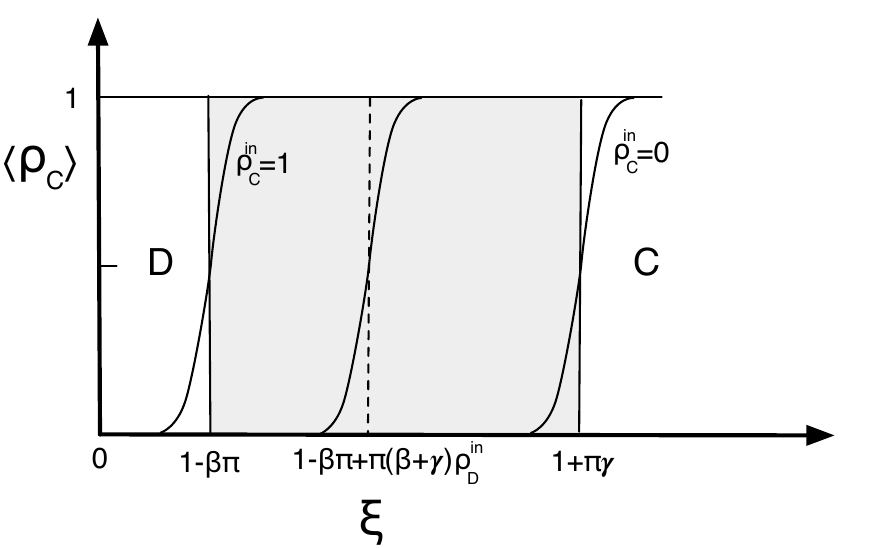} 
   \caption{Invasion probabilities for a fixed density of cooperators $\rho_C^{\rm in}$, as a function of the critical parameter $\xi$. Cooperation is stable for $\xi<1$ as long as the initial density of cooperators $\rho_C^{\rm in}$ is high. For $\rho_C^{\rm in}=1$, cooperators are stable at $\xi=1-\beta\pi$, which corresponds to the point $(k+1)(1-\beta\rho_P)$ in Fig.~\ref{dilemma2}. For general $\rho_C^{\rm in}$ the critical point is $\xi=1-\beta\pi+\pi(\beta+\gamma)(1-\rho_C^{\rm in})$, which may be larger or small than 1. For $\rho_C^{\rm in}$=0, the critical point is actually to the right of the critical point in the absence of punishment, that is, punishment hinders the establishment of cooperation. We sketch invasion probabilities as continuous across the critical lines to indicate the effect of finite population size. Increasing the population size creates steeper transitions approaching a sudden transition.}
   \label{dilemma3}
\end{figure}

What happens if a {\em group} of cooperators (rather than a single individual) tries to invade the defectors (or a group of defectors tries to invade the cooperators)? Because the fitness of any group is frequency-dependent, we have to recalculate the mean fitness of a group as follows: Assume a population of strategies given by the mean densities $(\rho_C,\rho_P)$. Let us also assume that, in general, cooperators punish with a probability $\pi_C$, while defectors punish with a probability $\pi_D$.  Then, $\rho_P=\pi_C\rho_C+\pi_D\rho_D$. The mean fitness of a group of cooperators is then given by
\be
\bar w_C=(1-\pi_C)P_C+\pi P_M=P_C-\gamma \pi_C \rho_D\;,
\ee
using the payoffs (\ref{eq1c}-\ref{eq4c}), and the fitness advantage of the cooperating type with respect to the defectors is
\be
\bar w_C-\bar w_D&=&P_C-P_D-(\pi_C-\pi_D)\gamma\rho_D\nonumber \\
&=&\xi-1+\beta(\pi_C+\pi_D)-(\pi_C-\pi_D)(\beta+\gamma)\rho_D\;.  \ \ \ 
\ee
We will see in the numerical results below that immoralists go extinct quickly (because they bear the double cost of meting out and receiving punishment). As a consequence, we set $\pi_D=0$ (defectors don't punish), and write the cooperator's probability to punish as $\pi_C\equiv \pi$, so that
\be
\bar w_C-\bar w_D=\xi-1+\beta\pi-\pi(\beta+\gamma)\rho_D\;. \label{def}\
\ee
Eq.~(\ref{def}) implies that punishment enables a ``premature" phase transition to cooperation as long as a ``nucleus" of cooperators $\rho_C^{\rm in}$ of sufficient size exists: a hallmark of metastability (see Fig.~\ref{dilemma3}). Thus, a ``fluctuation" of pure moralists ($\rho_C^{\rm in}=1, \pi=1$) is stable at $\xi=1-\beta$, which can be significantly smaller than 1 if the effect of punishment is large. However, the opposite dynamics occur for groups of defectors: they can invade stable cooperators at $\xi>1$ as long as the density of invading defectors $\rho_D^{\rm in}$ is large enough. As outlined in Fig.~\ref{dilemma3}, a ``fluctuation" into all defectors  $\rho_D^{\rm in}=1$ is stable for $\xi=1+\pi\gamma$, which can be substantially larger than 1 when the defectors displace perfect moralists ($\pi=1$). Thus, punishment enables both cooperation and defection in a meta-stable phase,  away from the critical point $\xi=1$. This behavior is akin to the phenomenon of supercooling/superheating, and can result in hysteresis: a population that starts in a defecting phase will stay in the cooperating phase past the critical point $\xi=1$ as $\xi$ is raised adiabatically from low values, and remain in the defecting phase past the critical point as $\xi$ is lowered from high values adiabatically. We will verify this behavior in the numerical simulations that follow.   

\section{Evolutionary simulation of Public Goods games} \label{sec-simul}
In this section we test the predictions of the (infinite population size) mean-field theory using agent-based simulations with finite population size. The population consists of 1,024 individuals who each have four (randomly assigned) opponents, that is, we use $k=4$ throughout in the results presented here (with some results for $k=8$). For populations of this size, neutral drift is negligible and results do not change qualitatively if populations are larger. However, the steepness of the transition between defection and cooperation may depend on the population size in the standard manner expected from finite-size scaling arguments (see, e.g.,\cite[p. 441]{Pathria1996}). 

\subsection{Game dynamics and Genetic Algorithm}
Since all opponents are also players, each individual plays $k+1$ games per update. The actual play of each individual is determined by their probabilities to cooperate $p_C$ and to punish $p_P$ encoded as two genetic loci, which can be thought of as the outcome of a network of genes that encode this decision. When mutating strategies, instead of mutating the individual genes that make up the decision pathway, we simply replace the parental probability $p_C$ by a uniformly drawn random number in the offspring. We will call the locus encoding the probability $p_C$ simply the ``C gene" and similarly for the punishment gene.

When every individual has played against its $k$ partners, 2 percent of the population is replaced using a Moran-like process~\cite{Moran1962} in a well-mixed fashion. The Moran-like process with a finite replacement rate interpolates between a true Moran process (replacement rate equals to inverse population size) and a Wright-Fisher process, where the entire population is replaced every update.
In our replacement scheme, the identity of the players in any group is unrelated to their ancestry so that, effectively, the members of a particular playing group are randomly selected from the population~\cite{Tayloretal2004}. With a replacement rate of 2\%, it takes on average 50 population updates until the entire population is replaced, that is, a single generation has elapsed. In our simulations, the fitness of each individual is cumulative, that is, the payoff obtained in the next update of the population is added to the payoff already obtained (until that player is removed).  However, we have tested that zeroing out the fitness after each update does not alter the game dynamics. 
We also verified that varying the replacement rate does not change the dynamics of the population in this game, unlike in the case where strategies communicate~\cite{Iliopoulosetal2010}. Indeed, if strategies make their play dependent on the last play, then replacing the opponent can introduce noise into the communication, resulting in different levels of cooperation. 

We verified that the probability for a player to encounter cooperators is independent of whether that player is a cooperator or a defector, as is required for well-mixed populations~\cite{FletcherDoebeli2009}. The accumulated payoff (fitness) is used to calculate the probability that this player's strategy will be chosen to replicate and fill the spot of a player that was removed in the Moran process. In case payoffs (calculated according to the equations above) are negative, we add a constant payoff to each and every strategy so that the relative payoffs are unchanged (it is known that such an offset does not alter the population dynamics). While the spatial version of the game shows somewhat different dynamics than studied here, we study the well-mixed version because it is amenable to theoretical prediction (see section~\ref{sec-theory}). 

The two genes of every individual mutate with a probability $\mu$ when replicated. As mentioned earlier, mutating a probability replaces the probability with a uniformly distributed random number. We used a fixed mutation probability ($\mu=2\%$ per locus) in the results presented here, except when we test the influence of the mutation rate on the broadening of the phase transition. We have previously studied the effect of varying mutation rate in this game~\cite{HintzeAdami2010} and found only a weak dependence on the location of the fixed point. 

\subsection{Line of Descent} \label{sec-LOD}
After 500,000 updates, the line of descent (LOD) of the population is reconstructed~\cite{Lenskietal2003,Ostmanetal2012}, by picking a random organism of the final population and following its ancestry all the way back to the starting organism. This is possible because no recombination occurs between genotypes: descent is entirely asexual. The LOD recapitulates the evolutionary dynamic of the population, because it contains the successive list of genotypes that have achieved fixation in the population. Because the population size is large, only a small fraction of mutations (on the order 1/$N$ where $N$ is the population size) find themselves on the LOD by chance. Thus, the LOD reflects the selective pressures operating on the population, and the fixed point of the evolutionary trajectory faithfully characterizes these pressures. The ancestral genotype that anchors all lines of descent is given by  the random strategy $p_C=0.5$ and $p_P=0.5$. Because there is only one species in these populations, the individual LODs of the population coalesce to a single LOD fairly rapidly (which is why it is sufficient to pick a random genotype for following the LOD). In other words, the common ancestor of the entire population is invariably fairly recent. To be certain that we deal with LODs that have coalesced when calculating strategy fixed points from the LOD, we routinely discard the last 50,000 updates (about 1,000 generations) from every run. When determining evolutionary fixed points for the trajectory, we also discard the first half, as the population trajectory may still be transient.

\section{Results}
\subsection{Evolutionary trajectories and fixed points}
As the strategies adapt to the environmental conditions (specified by the parameters that define the game, including the neighborhood size, the mutation rate, and the replacement rate), the probabilities change from their initial values $(p_C,p_P)=(0.5,0.5)$ towards the selected ``fixed point" strategy. In order to visualize the evolutionary trajectory of a population, we reconstruct the evolutionary line of descent of an experiment (LOD, see section~\ref{sec-LOD}), which tells the story of that adaptation, mutation by mutation. While the LOD in each particular run can show probabilities varying wildly, averaging many such LODs can tell us about the selective pressures the populations face. In particular, averaging the probabilities on the LODs after they have settled down, can tell us the {\em fixed point} of evolutionary adaptation~\cite{Iliopoulosetal2010}. We determine this fixed point by discarding the first 250,000 updates of every run (the transient), along with the last 50,000 (in order to remove the dependence of the LOD on the randomly chosen anchor genotype) and averaging the remaining 200,000 updates. Note that this fixed point is a computational fixed point only: we do not mean to imply that the population's genotypes all end up on this exact point. Rather, due to the nature of the game and the selective pressures that change as the composition of the population changes, the evolutionary trajectories approach this point and then fluctuate around or near it. Thus, the fixed point reflects the {\em mean} successful strategy given the conditions of the game. 
\begin{figure}[htbp] 
   \centering
\includegraphics[width=3in]{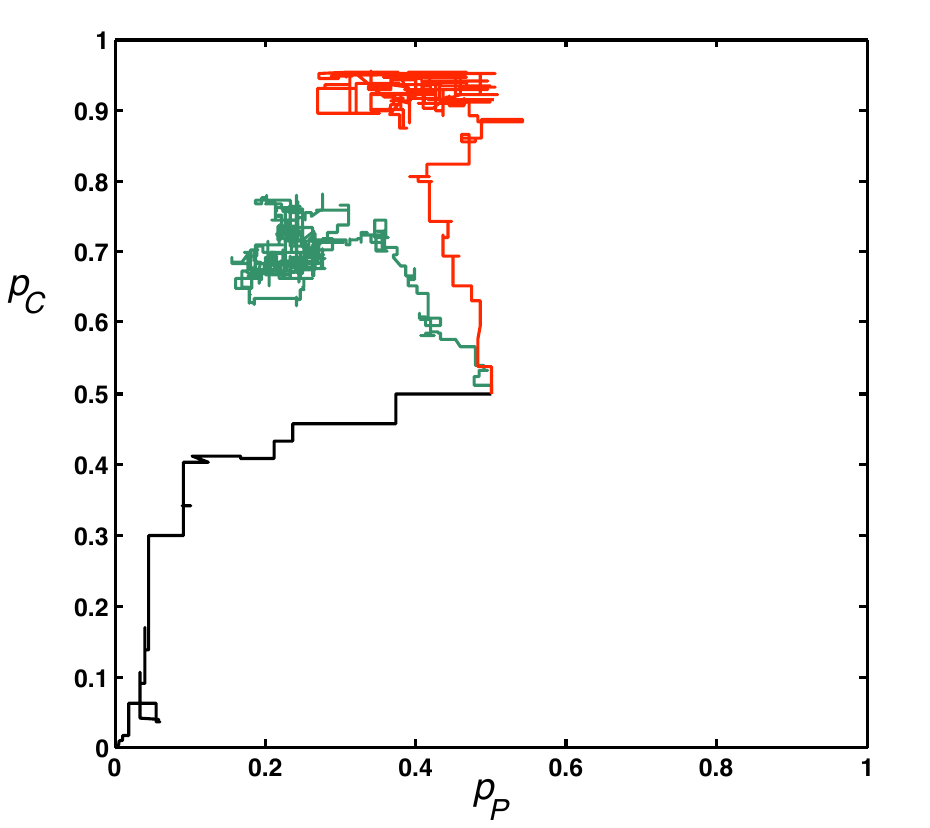} 
   \caption{{\bf Evolutionary trajectories for different synergies.} Evolution of strategies $(p_C,p_P)$ on the LOD for synergy factors $r=3$ (black), $r=4$ (green), and $r=5$ (red). All trajectories originate at (0.5,0.5). We show an average of the LOD of 10 runs each. Here, $\beta=0.8$, $\gamma=0.2$, and $\mu=2\%$.}
   \label{fig:traj}
\end{figure}

We show in Fig.~\ref{fig:traj} the average trajectories for three different synergy factors $r=3,4$, and 5 all anchored at the random strategy $(p_C,p_P)=(0.5,0.5)$ that was used as the seed strategy for every evolutionary run. We can see that, depending on the synergy (and the values chosen for the cost and effect of punishment), populations evolve towards a cooperating or defecting fixed point, and take different trajectories to get there. For $r=3$, synergy is too low to lead to cooperation, and the fixed point of that trajectory is $(p_C,p_P)=(0,0)$, that is, defection. For $r=4$, however, the population moves toward a fixed point centered around $(p_C,p_P)=(0.7,0.2)$, that is, players cooperate most of the time. (The location of the endpoint of the trajectory does not depend on the starting point.) Note, however, that the players engage in punishment only sparingly.  For $r=5$, cooperation is almost fully established, while punishment occurs about 40\% of the time on average. However, the average trajectory (average over ten independent runs) only tells part of the story, because at this level of cooperation there is very little difference between a punishing and a non-punishing player (given there are very few players to punish) and as a consequence the punishment gene has begun to drift. 
An unselected (and thus drifting) probability $p_P$ is a uniformly distributed random number, with mean 1/2 and variance 1/12. As $p_C\to1$, the average $p_P$ and its variance approach precisely these numbers. 

\begin{figure*}[!htbp]
\begin{center}
\includegraphics[width=6in,angle=0]{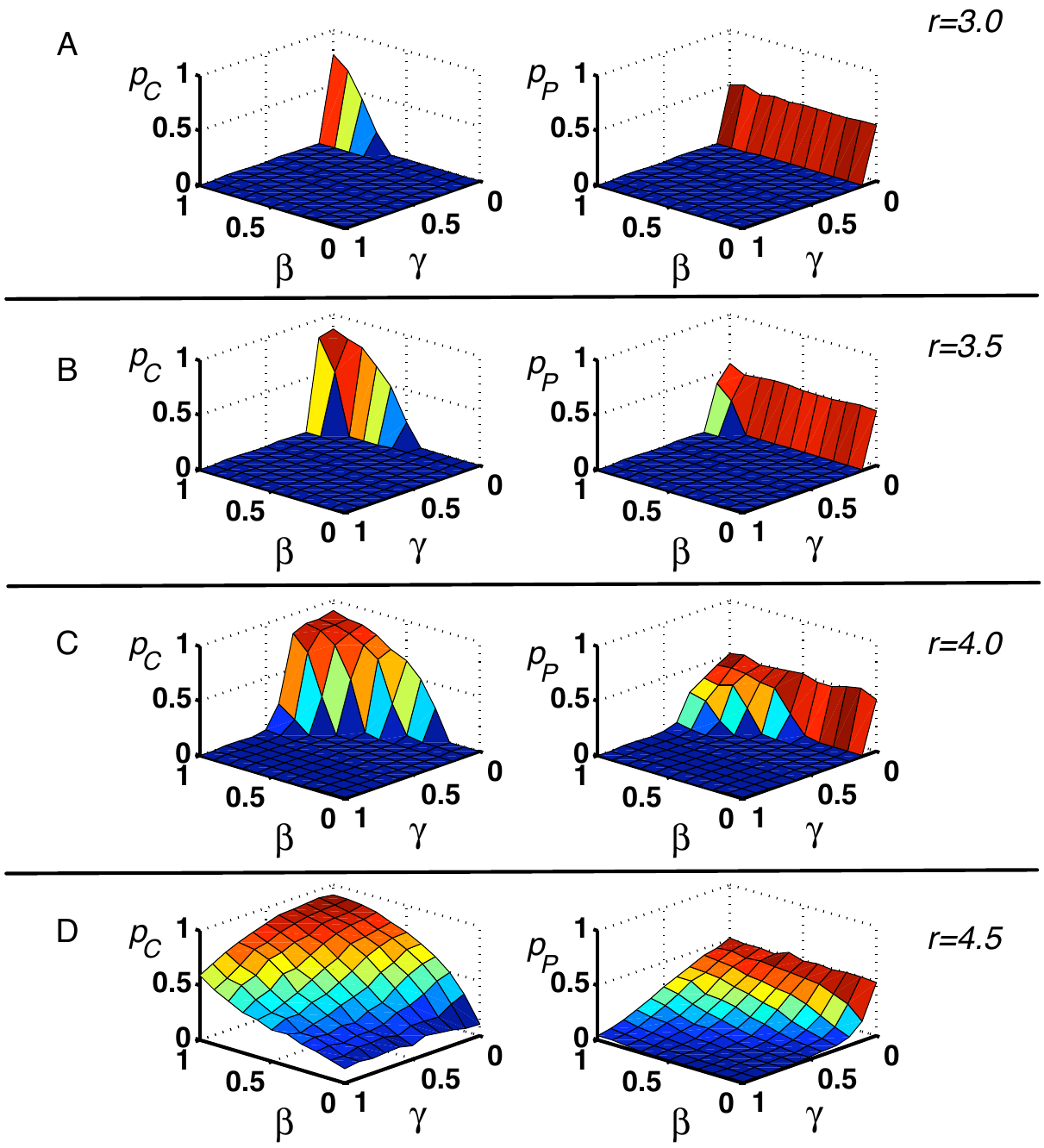}
\caption{{\bf Mean probabilities for cooperation $p_C$ and punishment $p_P$ at the evolutionary fixed point.} These graphs show the fixed point (averaged over 20 LODs) as a function of the cost of punishment $\gamma$ and the effectiveness of punishment $\beta$, for different values of the synergy $r$. Left panel: probability to cooperate $p_C$, right panel: probability to punish $p_P$. Note the inversion of the $\beta$ and $\gamma$ scales for better visibility. Mutation rate is set to $\mu=2\%$ per probability throughout. {\bf A}: For $r=3$, cooperation does not evolve except when punishment is free ($\gamma=0$), and even then only if punishment is very effective ($\beta$ close to 1). At $\gamma=0$, the punishment gene drifts neutrally.  {\bf B}: For $r=3.5$ defection is still the predominant strategy except for very low $\gamma$ and high $\beta$. {\bf C}: At $r=4$, cooperation is fully established for low $\gamma$ and high $\beta$, but not for medium values. {\bf D}: For $r=4.5$ cooperation is the dominant strategy for all values of the cost $\gamma$, and for high effect ($\beta>0.75$). Note that the average punishment probability $p_P$ never exceeds 0.5 (the value achieved when the gene drifts neutrally).}    
\label{fig-phase}
\end{center}
\end{figure*}

When mapping the strategy fixed point (average strategy on the LOD over 20 independent runs, again discarding the transient and the last 50,000) as a function of the parameters $\beta$ (effectiveness) and $\gamma$ (cost) of punishment (defined in section~\ref{sec-theory}) each in the range from 0.0 to 1.0 and at low synergy $r=3.0$, we find that defection is the most prevalent strategy on the LOD (see Figure~\ref{fig-phase}A), as was found previously~\cite{Brandtetal2003,Helbingetal2010}. When $\gamma=0$ there is no cost associated with the punishment, which implies that the P gene is not under selection and drifts.  Thus, for this value of synergy (and lower), we find that the strategy fixed point is defection without punishment, except for the values $\gamma=0$, where punishment is random.

As the degree of synergy increases to $r=3.5$, cooperation starts to appear even in this well-mixed population (see Fig.~\ref{fig-phase}B), while it appears as early as $r=2$ for sufficiently high $\beta$ and low $\gamma$ in the spatial (but deterministic) version of the game, see~\cite{Brandtetal2003,Helbingetal2010}. For $r=4$ we find players cooperating ($p_C\approx0.8$) at high $\beta$ and low $\gamma$ which indicates that under conditions where punishment is not very costly or even free, punishment pays off. In addition we notice that the probability to punish increases under the same conditions that allows cooperation (high $\beta$ and low $\gamma$, that is high impact, low cost of punishment), indicating that punishment is indeed used to enforce cooperation (Fig.~\ref{fig-phase}C). The mean punishment probability grows to 0.5, but at the same time the variance shows that this gene is not under selection (as long as $\gamma\neq0$). 

Increasing the synergy level even more towards $r=4.5$ we witness the emergence of dominance of cooperation ($p_C>0.5$) for most of the range of punishment cost and effectiveness, see Figure~\ref{fig-phase}D. At the same time the punishment probability reaches 0.5 for a larger range of parameters, but the mean punishment probability on the LOD never exceeds 0.5, implying that full persistent punishment is not stable, and probably not necessary. Note that, in an implementation where decisions are deterministic (such as in the implementation of Helbing et al.~\cite{Helbingetal2010}), punishment may remain for a long time in the population even though it is not selected anymore. In that case, players that cooperate with and without punishment have exactly the same fitness, and one or the other strategy should only dominate by drifting to fixation neutrally, a process that can take a significant amount of time in large populations such as those studied in Ref.~\cite{Helbingetal2010}.

\subsection*{Critical dynamics and the role of punishment}
\begin{figure}[htbp]
\begin{center}
\includegraphics[width=4in,angle=0]{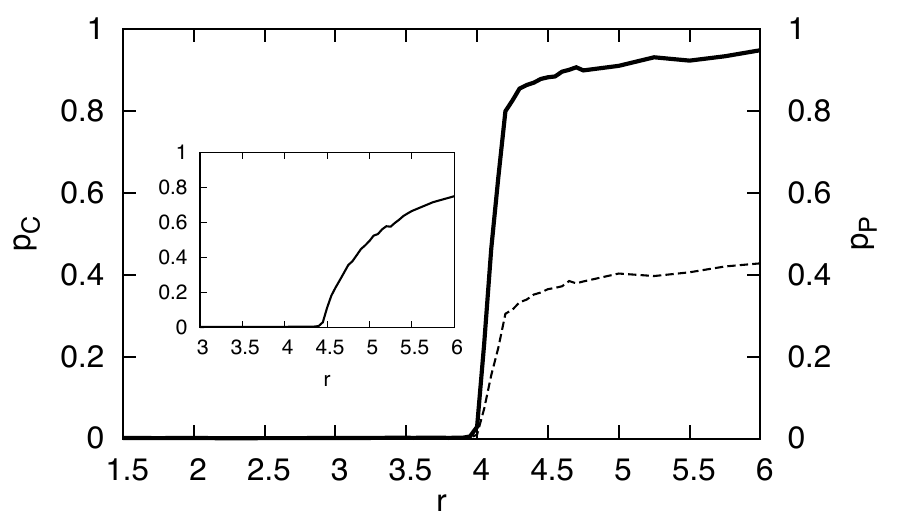}
\caption{{\bf Mean probability of cooperation and punishment}. Probability of cooperation $p_C$ (solid, left scale) and probability of punishment $p_P$ (dashed, right scale) with adaptive punishment at the evolutionary fixed point of the trajectory, as a function of the synergy $r$ ($\beta=0.8, \gamma=0.2, \mu=2\%$, 100 replicates for each data point). The probability to cooperate when punishment is forced to zero ($p_P=0$) is shown in the inset, for two different mutation rates: ($\mu=2\%$ (dashed) and $\mu=1\%$ (solid). }
\label{fig-crit}
\end{center}
\end{figure}

Previously, a phase transition between cooperative and defective behavior in the Public Goods game as a function of the synergy $r$ was observed for the spatial version~\cite{SzaboHauert2002,Brandtetal2003,Helbingetal2010c} of the game (but not the well-mixed version). We can study the critical point and its dependence on punishment in detail in the well-mixed version of the game, where analytical predictions (as outlined above) are available. We show in Fig.~\ref{fig-crit} the average probability to cooperate (solid line) and to punish (dashed line) as a function of synergy for our default values $\gamma=0.2$ and $\beta=0.8$. Cooperation sets in at $r=4$ and becomes prevalent for synergies just exceeding that. 
The simulations show that the phase transition is broadened from the expected first-order behavior, owing to two factors: finite population size and finite mutation rate. The inset in Fig.~\ref{fig-crit} shows the transition in the absence of punishment at two different mutation rates, suggesting that mutations introduce ``disorder-broadening"~\cite{ImryWortis1979}, which can lead to complete ``rounding" of the discontinuous transition. 

\begin{figure}[htbp]
\begin{center}
\includegraphics[width=4in,angle=0]{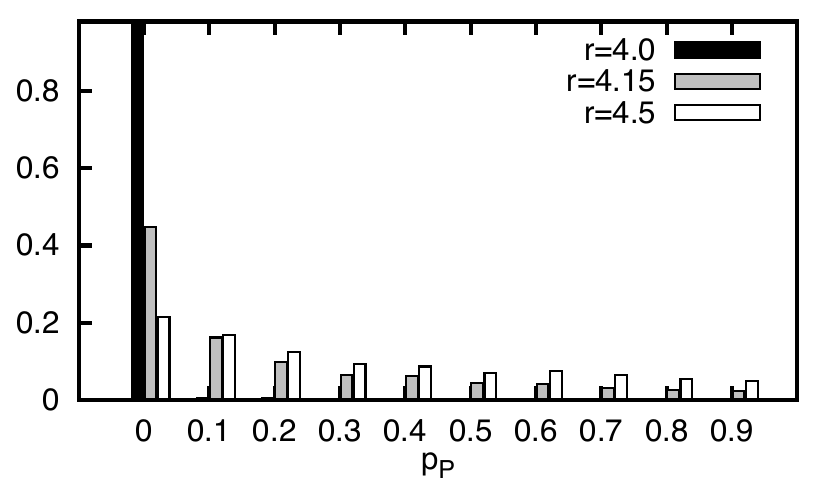}
\caption{{\bf Histogram of the punishment probability distribution.} Punishment probability distribution in a typical equilibrated population, just before the critical point ($r=4$, black), at the critical point ($r=4.15$, grey), and above $r_{\rm crit}$ ($r=4.5$, white).}
\label{fig-hist}
\end{center}
\end{figure}

We will now study how punishment affects the critical point.  The average probability of cooperation in Fig.~\ref{fig-crit} shows the typical behavior of an order parameter as a function of the critical parameter $r$ in a broadened first-order transition.  Thus, although punishment is sporadic when it is possible--and drifts when cooperation is established--it is essential to lower the critical barrier for cooperation. The probability distribution of the punishment gene throughout the population (Fig.~\ref{fig-hist}) shows that punishment is never prevalent: it is absent below the critical point, while the distribution is close to uniform (because of drift) above it. In a sense, punishment catalyzes the transition from defection to cooperation. Note also that the levels of cooperation achieved (at a given $r$) are significantly higher when punishment exists, even though punishment is only weakly selected for. Apparently, the possibility of punishment alone is sufficient to enforce higher levels of cooperation, but the mechanism for this enforcement is not immediately clear as punishment is rare above the critical point. 
 
In section~\ref{sec-theory} we calculated approximately the point at which cooperation is favored in a mean-field approach that does not take mutations into account,  by writing Eqs.~(\ref{eq1}-\ref{eq2}) in terms of the density of cooperators $\rho_C$ encountered by players in a group, and found that cooperation was favored as long as
\be
r>(k+1)(1-\beta\rho_P)\;.\label{crit}
\ee
This equation (which also follows from a replicator equation approach for the corresponding deterministic game) implies that the emergence of cooperation depends crucially on the density of punishers. In fact, the mean-field theory predicts that cooperation in the absence of punishment is favored only at $r=5$. We see cooperation emerge quite a bit earlier than that in our simulations (see inset in Fig.~\ref{fig-crit}), but crosses $p_C=0.5$ very close to $r=5$, as predicted by the mean field theory. Of course, the departure from the mean-field theory results is a consequence of the finite population size and mutation rate of the simulations. We also note that while the simulations suggest a stable fixed point $p_C<1$ above the critical point, this does not mean that the corresponding fixed point for the deterministic game admits a stable mixture of cooperating and defecting strategies. While from general arguments the fixed point of the stochastic game must be given by the corresponding fixed point of the deterministic game~\cite{Zeeman1981} (see also~\cite{Adamietal2012}), the ``broadening" of the phase transition via mutations (the equivalent of random quenched impurities~\cite{ImryWortis1979}) leads to a stochastic fixed point with $p_C<1$. But note that this fixed point is statistical only: for each run the trajectory fluctuates around this point.

\begin{figure}[!htbp]
\begin{center}
\includegraphics[width=5.25in,angle=0]{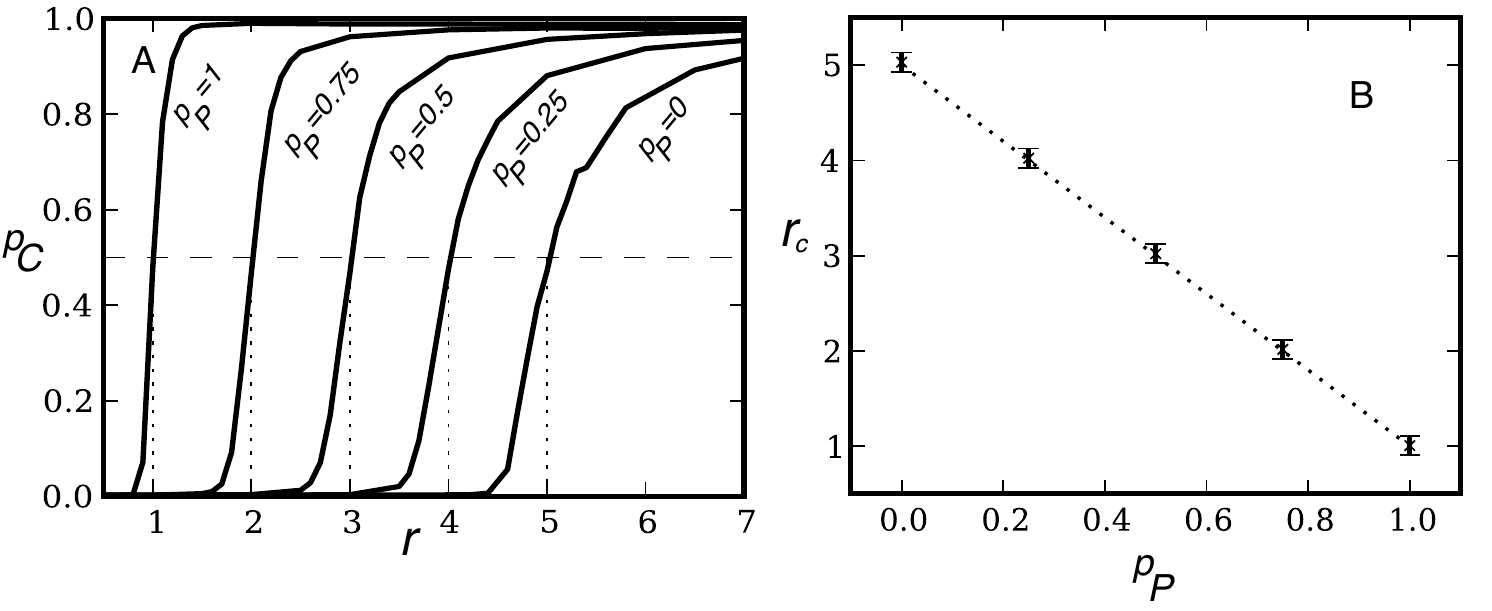}
\caption{{\bf Critical point at fixed punishment for $k=4$.} A: Mean probability to cooperate averaged over 100 independent lines of descent (average over 200K updates, discarding the first 250K and the last 50K as described in section~\ref{sec-LOD}, as a function of synergy $r$ for fixed (unevolvable) probability of punishment $p_P$=0.0, 0.25, 0.5, 0.75, and 1.0. The dashed line indicates a mean probability to cooperate $p_C=0.5$, which we us to extrapolate the critical value $r_c$. This critical value depends on the punishment levels as predicted by Eq.~(\ref{crit}).   B: Critical synergy $r_c$ as a function of punishment probability $p_P$ as deduced from panel A (points) by identifying the $r_{\rm crit}$ at which the cooperation probability $p_C=0.5$. The dashed line indicates the prediction $r_c=5(1-\beta\rho_P)$, assuming that the density of cooperators $\rho_P\approx p_P$ (mean field), with $\beta=0.8$ and $\gamma=0.2$. 
}
\label{fig-pred}
\end{center}
\end{figure}
We can test Eq.~(\ref{crit}) explicitly by finding the critical $r$ at which $p_C$ crosses 0.5 for simulations in which the punishment probability is held fixed, so that $\rho_P \approx p_P$. To find the critical point, we performed 100 simulations each at fixed $r$ with small increments $\Delta r$ and interpolated the data within the steep portion of the transition to find the crossover point. The curves in Fig.~\ref{fig-pred} show that the steepness of the transition between cooperation and defection depends on the level of punishment, changing from a dependence reminiscent of second-order transitions (at vanishing punishment) towards a first-order-like transition at high punishment. We plot the critical line $r_c=(k+1)(1-\beta\rho_P)$ in Fig.~\ref{fig-pred} for $k=4$ and $\beta=0.8$ ($r_c=5-4p_P$). The mean field theory reproduces the simulated $r_c$ within errors. The prediction in fact works just as well for other parameter values: we tested $k=8$ (each agent plays with eight random other agents) and readily observe that the critical value is given by $r_c=10-8p_P$ (data not shown).  

Because of the crucial importance of punishers in determining the synergy level at which cooperation emerges, the Public Goods game with a genetic basis (that is, with genes coding for probabilities of moves) implies curious dynamics close to the critical point. Below the critical point, defection is a stable strategy, and punishment is absent. When cooperation emerges as a possibility, punishment becomes more and more important, leading to a {\em lowering} of the critical synergy for cooperation via Eq.~(\ref{crit}). At that point, cooperation emerges rapidly and decisively once a critical level of punishment has been achieved.
Once cooperation is dominant and defectors are all but driven to extinction, punishment becomes irrelevant and the gene for punishment begins to drift. As this happens, the fraction of punishers drops, thus raising the critical synergy according to Eq.~(\ref{crit}). As a consequence, a drifting punishment gene can lead to the sudden re-emergence of defectors as stable states. Once those have taken over, the reverse dynamics begins to unfold. Given this dynamic, we should observe periods of cooperation and defection that follow each other closely when the synergy is near the critical point. 

These dynamics are reminiscent of the phenomenon of supercooling and superheating in certain phase transitions observed in condensed matter physics, as predicted in section~\ref{sec-theory}. If we imagine the synergy parameter $r$ as the critical parameter and the mean probability to cooperate as the order parameter, it is possible that when $r$ is slowly increased, the population remains in the defecting phase because a switch to cooperation requires a critical number of cooperators as a ``seed". In such a situation, the defecting phase is unstable to fluctuations. If a critical number of cooperators emerges by chance, punishment immediately becomes effective against defectors, lowers the critical point as implied by Eq.~(\ref{crit}), and the population could transition to cooperation very quickly. A hallmark of such bi-stable systems that require nucleation events in order to transition is {\em hysteresis}, a phenomenon where the state of the system depends on its history. We can test whether hysteresis exists in the Public Goods game (and whether the strength of this effect depends on the probability to punish), by adiabatically changing the synergy parameter first from low to high (transitioning from defection to cooperation), and then adiabatically back from high to low. 
\begin{figure*}[!ht]
\begin{center}
\includegraphics[width=6in,angle=0]{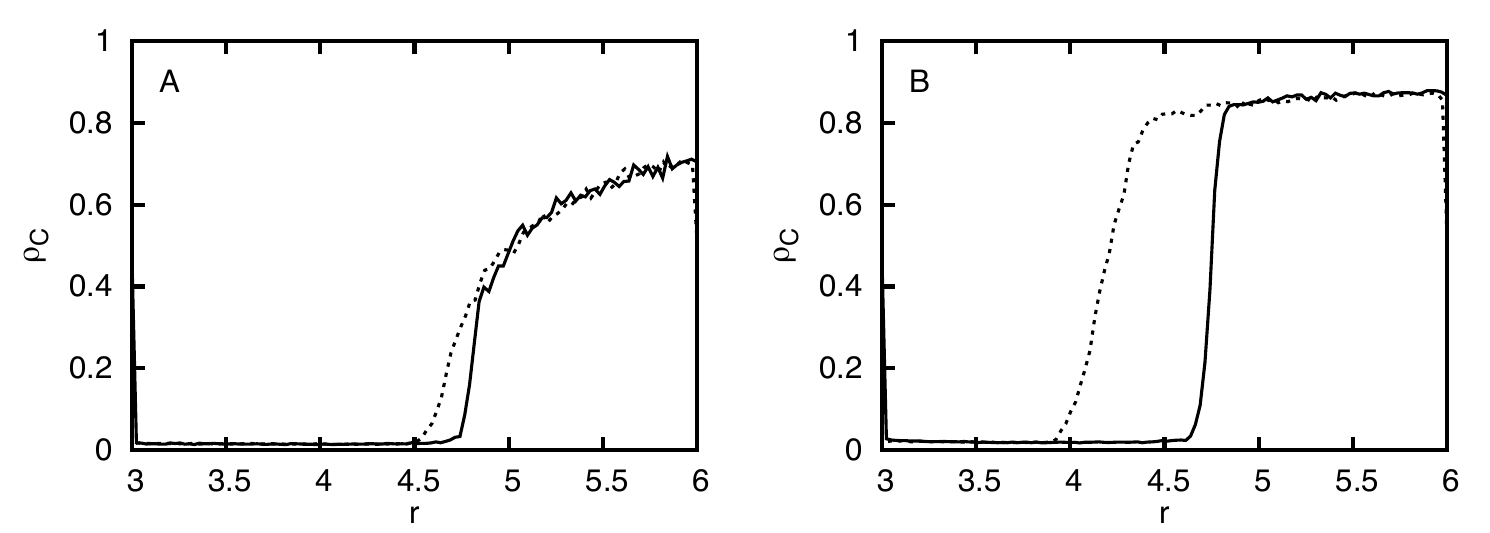}
\caption{{\bf Hysteresis effect from punishment.} Population fraction of cooperators (measured as the density of non-punishing cooperators plus the density of moralists) as a function of synergy $r$ when $r$ is adiabatically changed from low to high values (solid), and back from high values to low values (dashed). A: no punishment. B: Adaptive punishment with $\beta=0.8$ and $\gamma=0.2$. All population fractions are started at $0.5$ (either at the high or low end of $r$). The lines show the average over 100 runs. Standard error is of the size of the fluctuations.}
\label{fig-hyst}
\end{center}
\end{figure*}
While we see evidence of hysteresis even when punishment is absent (Fig.~\ref{fig-hyst}A), the effect is much more pronounced when punishment is possible (Fig.~\ref{fig-hyst}B). The population moves from cooperation to defection at about the expected critical synergy $r_{\rm crit}\approx 4.15$ as $r$ is decreased, but stays in the defecting phase much beyond the critical point as $r$ is increased.  

The observed hysteresis effect implies that once cooperation is established, it can be maintained even when the expected synergy fluctuates below the critical point, but that cooperation is difficult to establish even when the synergy would be conducive for that establishment. It also explains why levels of cooperation are higher when punishment is possible, even if punishment is used sparingly. In meta-stable phase transitions, bubbles of the new phase increase in size exponentially if larger than a critical size, but shrink exponentially when smaller than the critical size~\cite{LandauLifschitz1987}. Thus, if a group invades with $\rho_C^{\rm in}>\rho_{\rm crit}$, $\rho_C\to 1$. This is different from the 
dynamics in the absence of punishment, where at the critical point all $\rho_C^{\rm in}$ have the same fitness, and the mean level of cooperation is 0.5, as is evident in Fig.~4 (inset), and Fig. 8A. Indeed, the critical point for $\rho_P=0$ (no punishment) is neutral, while it is a repulsive fixed point when punishment is present. As a consequence, the phase transition as a function of $r$ becomes steeper and steeper as punishment increases, and higher levels of cooperation are achieved. This behavior is strongly reminiscent of phase transitions in ferromagnetic systems where the ``rounding" of the transition due to impurities is reduced via the magnetic field. This suggests that a treatment of Public Goods games in terms of Ising-like models where punishment plays the role of a magnetic field forcing the alignment of spins should be possible, and we are currently pursuing such an approach~\cite{Adamietal2015}.

\section{Discussion}
We studied the Public Goods game for well-mixed populations both theoretically and in agent-based simulations of 
Darwinian evolution of stochastic strategies, using genes that encode the probabilities for cooperation and punishment. It is known that punishment can drive the evolution of cooperation above a critical synergy level as long as there is a spatial structure in the environment~\cite{Brandtetal2003,Helbingetal2010}. It was also previously believed that in well-mixed populations cooperation via punishment can only become successful if additional factors like reputation~\cite{Sigmundetal2001} or the potential for abstaining from the public good~\cite{SzaboHauert2002,Hauertetal2007} are influencing the evolution. Here we show that cooperation readily emerges in a well-mixed environment above a critical level of synergy. This critical level is influenced by a number of factors: the rate of punishment because punishment favors cooperating groups, but also spatial structure~\cite{SzaboHauert2002,Brandtetal2003,Helbingetal2010c}, because a single cooperator can nucleate a transition simply because offspring cooperators are placed next to it, giving rise to a ``bubble" of cooperators of sufficient size. This finding is similar to the observation of frequency-dependent cooperation when punishment (but not cooperation) is probabilistic~\cite{Chenetal2014}, even though the game studied by Chen et al.\ is different from the game we consider here in that punishment is shared among the punisher, whereas in our game (which is the one studied by Helbing et al.~\cite{Helbingetal2010}), each punisher acts on his own. 

We have not studied here the possibility of ``anti-social" punishment\cite{NakamaruIwasa2006,Randetal2010,RandNowak2011}, where non-cooperating defectors can punish cooperators, but we do not expect this possibility to change the overall picture. Indeed, in simulations in which defection was not punished but instead rewarded (a negative punishment), this only served to reinforce the defecting phase, as was also found in~\cite{Hauseretal2014}. A transition to the cooperative phase still takes place at sufficiently high synergy. Phase transitions between cooperative and defection phases have also been observed in a spatial version of the Public Goods game where costly rewards are given for cooperation, rather than the costly punishment for defectors~\cite{SzolnokiPerc2012}. It would be interesting to study this game within the context of evolving probabilistic strategies. 

We conclude that in well-mixed populations cooperation can emerge if the synergy outweighs the defectors' reward, which is reduced by punishment. A punishment-dependent barrier to cooperation introduces an interesting dynamic near the critical synergy. Starting in the cooperative phase, as long as the mutation rate is low enough, the dearth of defectors in the cooperating phase makes punishment obsolete, that is, the selective pressure to punish disappears. As a consequence, the density of punishers decreases, thus increasing the critical point in turn. If the critical synergy has increased sufficiently, defectors can again gain a foothold. Such a shift, however, reinstates the selective pressure to punish, leading to a re-emergence of moralists that can drive defectors out once more. Thus, for synergy factors near the critical point, we can expect oscillations between cooperators and defectors, and no strategy is ever stable. 

Finally, the observation of hysteresis  implies the existence of metastable states, and gives rise to a self-enforcing (or ``self-aligning") dynamic where cooperation is stable even when punishment is never actually used. It is clear that meta-stable dynamics can only occur in the shaded region in Fig.~3, which is set by the probability $\pi$ with which cooperators punish, and provides a mechanism to protect cooperating groups from defectors, as the defectors need to achieve a critical density in order to thrive. In a very real way, meta-stability raises the scepter of punishment to maintain cooperation, even when it is not used. 


\section*{Acknowledgements}We acknowledge useful discussions with M. Mirmomeni, and substantive comments by two reviewers. This work was supported in part by NSF's BEACON Center for the Study of Evolution in Action, under Contract No. DBI-0939454.  We wish to acknowledge the support of the Michigan State University High Performance Computing Center and the Institute for Cyber-Enabled Research.

\section*{Bibliography}


\end{document}